\documentclass[twocolumn, preprintnumbers, amsmath, amssymb] {revtex4}
\usepackage{graphicx}
\usepackage[usenames]{color}
\usepackage{dcolumn}
\usepackage{bm}

\newcommand{\be}{\begin{equation}}
\newcommand{\ee}{\end{equation}}

\begin{document}

\title{Generality of shear thickening in dense suspensions}
\author{Eric Brown$^{1}$}
\email{embrown@uchicago.edu}
\author{Nicole A. Forman$^{2,3}$}
\author{Carlos S. Orellana$^1$}
\author{Hanjun Zhang$^3$}
\author{Benjamin W. Maynor$^2$}
\author{Douglas E. Betts$^3$}
\author{Joseph M. DeSimone$^{2,3}$}
\author{Heinrich M. Jaeger$^1$}
\affiliation{$^1$James Franck Institute, The University of Chicago, Chicago, IL 60637\\
$^2$Liquidia Technologies, Research Triangle Park, NC 27709\\
$^3$Department of Chemistry, University of North Carolina, Chapel Hill, NC 27599}

\date{\today}


\maketitle

{\bf Suspensions are of wide interest and form the basis for many smart fluids \cite{St04, WHYLS03, TPCSW01, JBC98, SWB03, LWW03, ZWG08}.  For most suspensions, the viscosity decreases with increasing shear rate, i.e.~they shear thin. Few are reported to do the opposite, i.e.~shear thicken, despite the longstanding expectation that shear thickening is a generic type of suspension behavior \cite{BB85, Ba89}.  Here we resolve this apparent contradiction.  We demonstrate that shear thickening can be masked by a yield stress and can be recovered when the yield stress is decreased below a threshold.   We show the generality of this argument and quantify the threshold in rheology experiments where we control yield stresses arising from a variety of sources, such as attractions from particle surface interactions, induced dipoles from applied electric and magnetic fields, as well as confinement of hard particles at high packing fractions.   These findings open up possibilities for the design of smart suspensions that combine shear thickening with electro- or magnetorheological response.}

Shear thickening is presumed to be due to general mechanisms such as hydrodynamics \cite{BB85, MW01a} or dilation \cite{Ho82, LDHH05, FHBOB08}, and thus all suspensions are expected to exhibit shear thickening under the right conditions \cite{Ba89}. So far, however, the exact conditions have not been determined.  One condition is apparently set by attractive particle interactions. It has long been known that attractions, observed as flocculation in suspensions, can prevent shear thickening.  This has been shown by modifying the chemistry, for example by adding flocculating agents to observe the transition from shear thickening to thinning (for a review, see\cite{Ba89}).  In other cases, crossing the gel transition was shown to eliminate shear thickening \cite{GP43, GZ04}.  A key problem, therefore, is to understand how interparticle attractions interfere with shear thickening.  We demonstrate here that a simple and direct criterion for the existence of an observable shear thickening regime in dense, non-Brownian suspensions can be developed by comparing  the yield stress produced by attractions with the inherent shear thickening stresses.   We then generalize this condition to show how a yield stress from any source modifies the shear thickening phase diagram.  

\begin{figure}                                                
\centerline{\includegraphics[width=3.in]{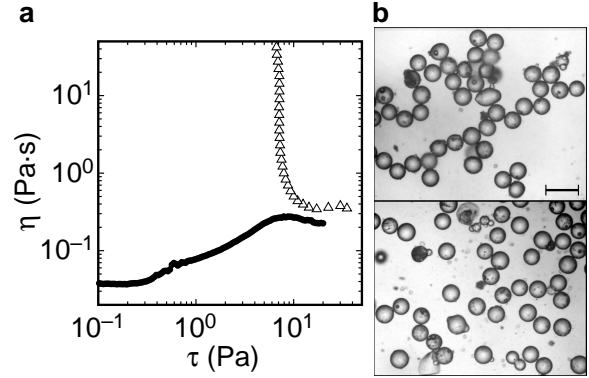}}
\caption{{\bf Revealing shear thickening by adding surfactant to hydrophobic glass spheres in water.}  Soda lime glass spheres of mean diameter 90 $\mu$m  with a hydrophobic silane coating were suspended at a packing fraction $\phi=0.52$.  {\bf a}, $\bigtriangleup$:  viscosity curve without surfactant.  The divergence of the curve is characteristic of a yield stress.  $\bullet$:  viscosity curve of the same system at the same $\phi$ with added surfactant.  The shear thickening regime is the region of positive slope in the curves of viscosity $\eta$ versus applied stress $\tau$.  Shear thinning is characterized by a negative slope and Newtonian fluids, such as water, exhibit  constant $\eta$.  {\bf b}, Images show clustering due to interparticle attractions (top) and no clustering when surfactant is added (bottom).  Scale bar is 200 $\mu$m.  All images (including subsequent figures) were taken at rest under an optical microscope in a dilute quasi two-dimensional layer.  In this dilute case, attractions can be observed by the high number of particle contacts in the form of clusters or chains. }  
\label{fig:hydrophobic}                                        
\end{figure}

Our experiments used an Anton Paar rheometer to measure the shear stress $\tau$ and the shear rate $\dot\gamma$ of a wide range of different suspensions.  The viscosity is defined as $\eta\equiv\tau/\dot\gamma$ in the steady state. Our focus is on non-Brownian, dense suspensions that show strongly packing-fraction-dependent, reversible shear thickening, often called `discontinuous,' because of the steep stress-shear rate relationship.  To understand the significance of interparticle attractions, we first consider the particle-liquid surface tension. Figure~\ref{fig:hydrophobic} shows the striking change in behavior produced by adding a small amount of surfactant to a water suspension of glass spheres with a hydrophobic coating.  In the aqueous environment the coating leads to network-like particle clusters (Fig.~\ref{fig:hydrophobic}b, top) which minimize exposed surface area and thus potential energy. As a consequence,  to pull particles apart requires overcoming a significant stress threshold.  In Fig.~\ref{fig:hydrophobic} this translates into a region where, for applied stresses smaller than this yield stress, the suspension does not flow and the viscosity effectively diverges.  The result is shear thinning behavior ($\bigtriangleup$).  Added surfactant eliminates the clustering with its associated yield stress and reveals a region of underlying shear thickening ($\bullet$) below the range of the previous yield stress.  This suggests the yield stress due to attractions is responsible for hiding shear thickening if it overwhelms the shear thickening stress range.   

\begin{figure*}                                                
\centerline{\includegraphics[width=7.25in]{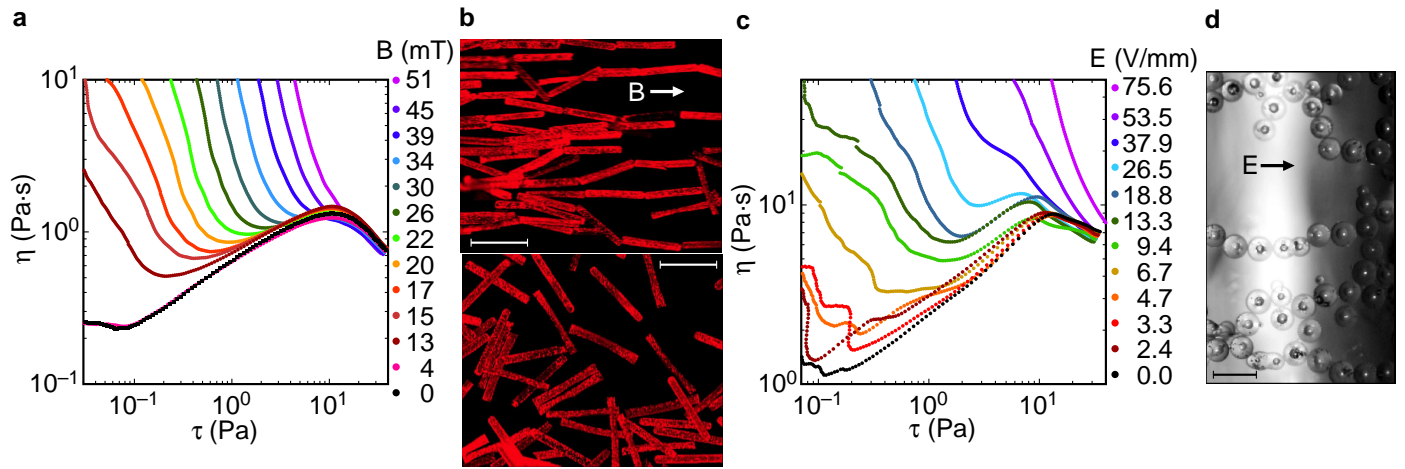}}
\caption{{\bf Using magnetic and electric fields to tune the interplay between shear thickening and the yield stress.} {\bf a}, Viscosity curves for a suspension of ferromagnetic rods ($254\times32\times25$ $\mu$m) for different values of applied magnetic field $B$.  Magnetite-doped (30\% by weight), PEG rods made by the PRINT process \cite{RMEEDD05} were suspended in PEG at a packing fraction $\phi= 0.20$.  The shear thickening region is seen to shrink and eventually becomes eliminated as it is encroached on by the increasing yield stress.  {\bf b},  Microscope images show the rods for $B=30$ mT (top) and $B=0$ (bottom).  {\bf c}, Viscosity curves for a suspension of dielectric spheres for different values of applied electric field $E$.  Soda-lime glass spheres of diameter 90 $\mu$m  were suspended in $58$ mPa$\cdot$s mineral oil at a $\phi=0.56$.  {\bf d}, The microscope image shows the spheres for $E=60$ V/mm.  At $E=0$, the image is similar to the bottom panel of Fig.~\ref{fig:hydrophobic}b.  In both panels b and d the fields were applied vertically, in the direction of the shear gradient in a parallel plate rheometer.  The scale bars are each 200 $\mu$m.
} 
\label{fig:STwfield}                                        
\end{figure*}

For a more detailed exploration than afforded by chemical means of the role played by the magnitude of the yield stress in modifying the shear thickening regime, we require in-situ, tunable control over the strength of the attractions.  This can be achieved by applied electric and magnetic fields that polarize particles of a given dielectric or magnetic susceptibility and also have the practical advantage of reversibility.  The result is a field-dependent attraction between neighboring particles and thus a continuously tunable yield stress.  We used dielectric glass spheres in mineral oil for electrorheology and magnetite-filled polyethylene glycol (PEG) rods suspended in PEG for magnetorheology.  Figure \ref{fig:STwfield} shows the evolution of the yield stress and shear thickening regime with both types of field.  For small fields, the viscosity curve is seemingly unaffected.  A main result from these data is that increasing the field strength, and the concomitant yield stress, pushes the onset  of shear thickening to higher stress values.   At intermediate field values the curves rejoin the zero-field shear thickening behavior after exhibiting a viscosity minimum.  A yield stress thus simply results in a smaller range of applied stresses over which shear thickening is observable.  Only when the yield stress becomes large enough that it encroaches on the upper limit of the shear thickening range is the effect fully eliminated.  Qualitatively this behavior is neither dependent on the suspension nor the source of the yield stress, as seen from the similarity between panels a and c in Fig.~\ref{fig:STwfield}.   The fact that the same conclusion can apply to Fig.~\ref{fig:hydrophobic} is especially remarkable considering that the induced dipoles are directional but the chemical attractions are anisotropic.  Given the different microstructures, this indicates that it is the stress scale resulting from attractions that determines whether shear thickening is observable or not.

\begin{figure}                                                
\centerline{\includegraphics[width=3.in]{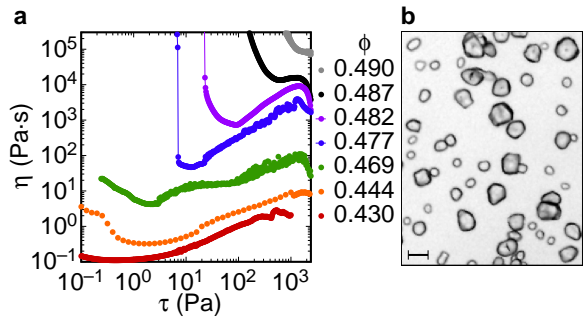}}
\caption{ {\bf Elimination of shear thickening by increasing packing fraction.} {\bf a}, Viscosity curves for cornstarch in water  at different packing fractions $\phi$.  The cornstarch particles had an average diameter of 14 $\mu$m.  The water was density-matched to 1.59 g/mL by dissolving CsCl in it.  A solvent trap was used to avoid evaporation and a Couette geometry was used to ensure the sample remained confined.  {\bf b}, The microscope image shows that particles do not cluster without confinement, also confirmed by optical tweezer measurements.  Scale bar is 20 $\mu$m.
}  
\label{fig:confinement}                                        
\end{figure}

The experiments discussed so far concerned yield stresses produced by particle attractions.  Similar behavior carries over to suspensions without attractive interactions in which a yield stress arises due to confinement at large packing fractions \cite{BJ09}.  Data are shown in Fig.~\ref{fig:confinement} for several different packing fractions of cornstarch in water.  It is seen again that the shear thickening range decreases as the yield stress increases and eventually disappears when this yield stress approaches the upper stress limit of the shear thickening regime.

\begin{figure*}                                                
\centerline{\includegraphics[width=6in]{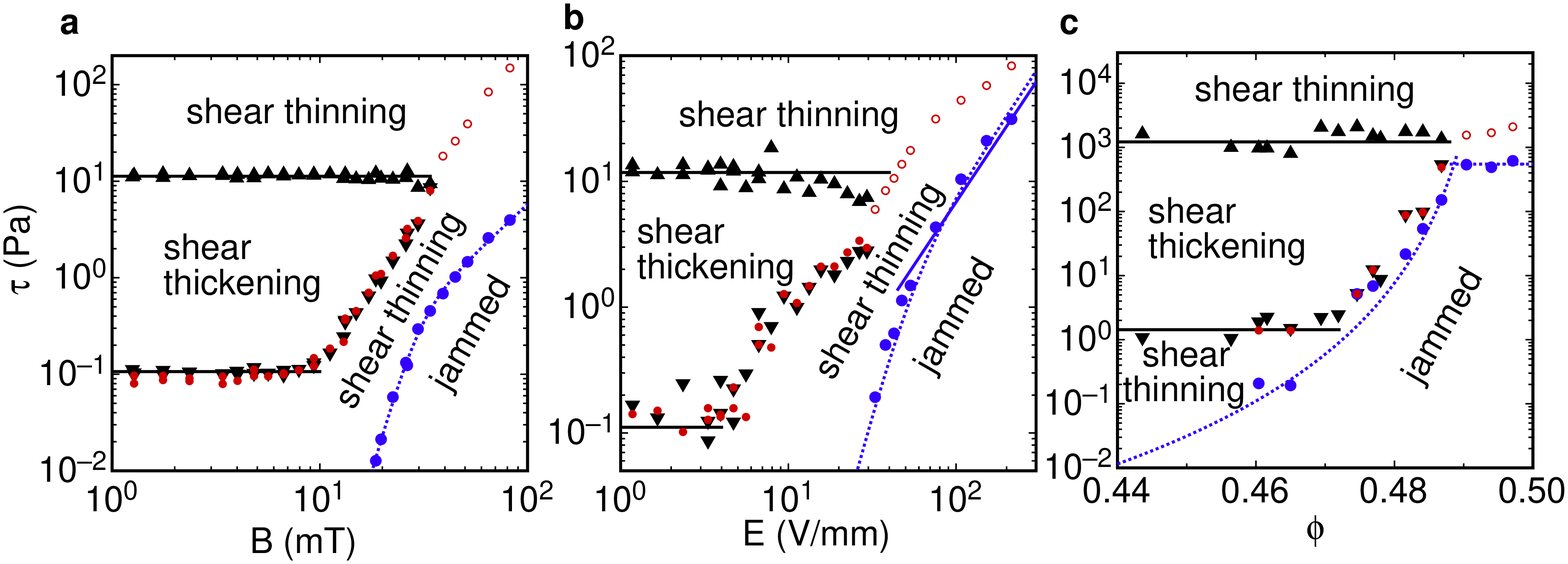}}
\caption{{\bf Non-equilibrium phase diagrams delineating observable shear thickening regions in terms of the associated stress range.} Stress range as a function of applied magnetic field $B$ (a), applied electric field $E$ (b), and packing fraction $\phi$ (c).   The boundaries of the shear thickening regime are set by the local minima ($\blacktriangledown$) and maxima ($\blacktriangle$) of the viscosity curves in Figs.~\ref{fig:STwfield} and \ref{fig:confinement}.  ({\color{Cyan} $\bullet$}): the yield stress $\tau_y$, below which suspensions are jammed.  ({\color{Red} $\bullet$}):  the predicted onset of shear thickening $\tau_m$ evaluated from Eq.~\ref{eqn:tau_min} at the measured $\dot\gamma_{m}$, demonstrating that the boundary is determined by the total shear thinning stress $\tau_{HB}$, regardless of the source of the yield stress.  For panels a and b the values of $\epsilon$ used is that measured for zero attractions, showing that the shear thickening stress term is independent of field strength.  For panel c the value of $\epsilon = 0$ is used which is measured at the highest packing fractions where shear thickening can be observed, showing that the phase boundary is equal to the shear thinning stress $\tau_{HB}$ in the limit of $\epsilon=0$.  Solid black lines:  the measured stresses at the upper and lower phase boundaries in the limit of zero field and small $\phi$.  These coincide with the measured phase boundaries for $B=0$ and $E=0$.  Solid blue line: prediction of the ER yield stress from two-particle interaction (see Suppl. Mat.). Dotted blue lines:  guides to the eye for the phase boundary between shear thinning and jammed regimes.  ({\color{Red} $\circ$}):  predicted $\tau_m$ in cases where no shear thickening regime was found using model predictions for $\dot\gamma_m$.  In each case, these values are close to or above the upper stress boundary, showing that the reason shear thickening was not found was because $\tau_m$ exceeded the shear thickening stress range.
}
\label{fig:phasediagrams}                                        
\end{figure*}

The interplay between yield stress and shear thickening emerging from the  data in Figs.~\ref{fig:STwfield} and \ref{fig:confinement} can be summarized in a set of non-equilibrium phase diagrams (Fig.~\ref{fig:phasediagrams}) showing the shear thickening, shear thinning, and jammed (defined here as a non-flowing state below the yield stress) regimes.  Despite the differences in sources of a yield stress, there are important similarities.  The stress thresholds bounding this regime (horizontal black lines) are nearly independent of $\phi$ when the yield stress is small enough \cite{MW01a, EW05, BJ09}.  As the yield stress increases, the lower threshold moves upward and eventually approaches the upper boundary, at which point shear thickening ceases.  For intermediate values of $B$, $E$, or $\phi$, both jamming and shear thickening can be found at different stress values \cite{HAC01, SK05, FHBOB08}.

Since the boundaries of the shear thickening region are determined by local extrema of viscosity curves, they can be calculated given the relation between stress and shear rate in the lower shear thinning and shear thickening regimes. Note that the yield stress value is below the shear thickening phase boundary, leaving a shear thinning regime between the jammed and shear thickening regions.  To quantify the effect of the yield stress on the shear thickening phase boundary, we therefore must account for this additional regime.   To model these contributions, we use the Herschel-Bulkley form, with a fixed exponent of $1/2$ commonly used to describe shear thinning behavior \cite{MW01a}

\be
\tau_{HB}(\dot\gamma) = \tau_y+a_1\dot\gamma^{1/2} \ .
\label{eqn:HBmodel}
\ee

\begin{figure}                                                
\centerline{\includegraphics[width=2.in]{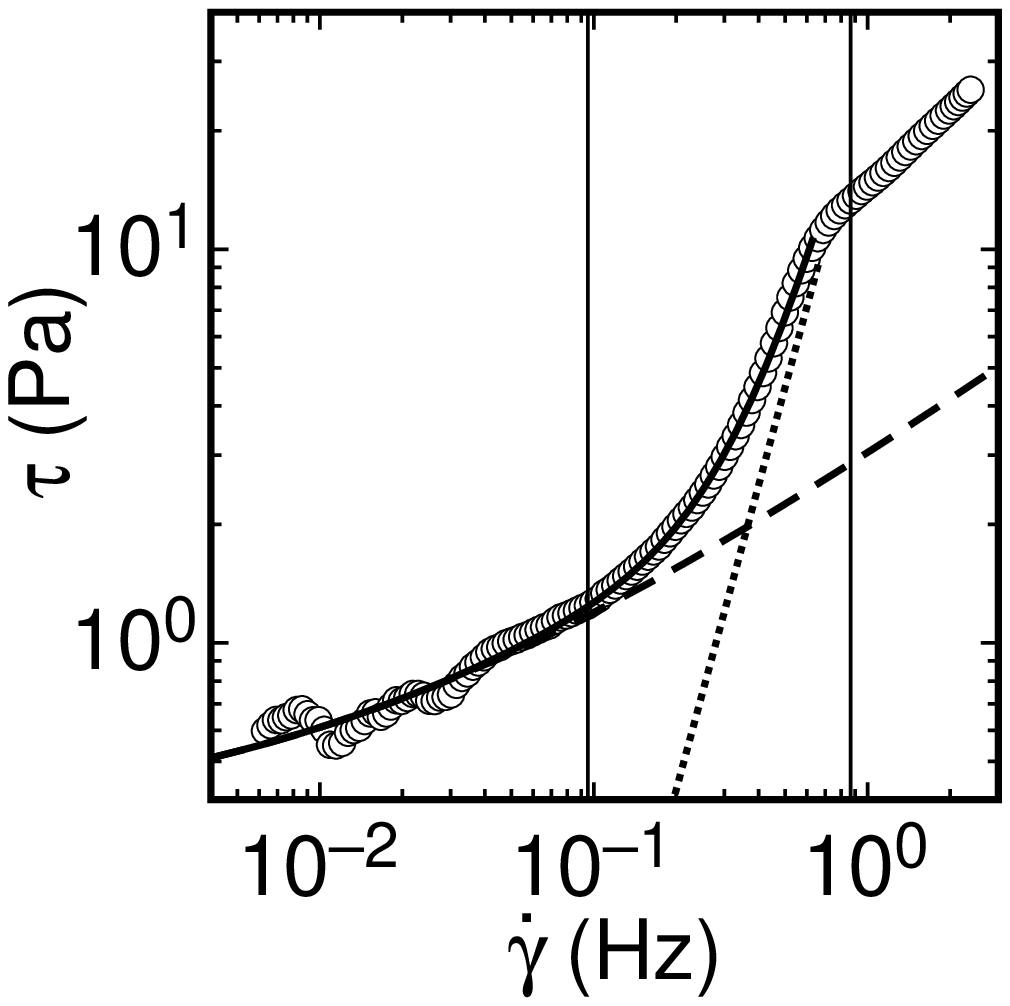}}
\caption{ {\bf Fit of a stress/shear-rate curve broken up into shear thinning and thickening components.} Data shown for glass spheres at $E = 12.5$ V/mm.   Dashed line:  fit of Herschel-Bulkley model (Eq.~\ref{eqn:HBmodel}) to the lower shear thinning regime.   Dotted line: term $\sim \dot\gamma^{1/\epsilon}$ representing the shear thickening regime.  Solid line:  sum of dashed and dotted lines (Eq.~\ref{eqn:HBmodel_plus}).  Vertical lines:  shear rates corresponding to the boundaries of the shear thickening range in the limit of $E=0$.  
}  
\label{fig:HBfit}                                        
\end{figure}

\noindent   Here the first term $\tau_y$ denotes the yield stress and $a_1$ parameterizes the additional stress that is operative in the shear thinning regime.  We refer to $\tau_{HB}$ as the shear thinning stress.  Earlier work  \cite{BBV02, MB04a, GZ04} suggests that contributions to the overall shear stress can be linearly separated as

\be
\tau(\dot\gamma) = \tau_{HB}(\dot\gamma) + a_2\dot\gamma^{1/\epsilon} \ ,
\label{eqn:HBmodel_plus}
\ee

\noindent where the second term represents the shear thickening stress parameterized by  $a_2$ and an exponent $\epsilon$ that approaches zero in the limit where the stress/shear-rate relation becomes discontinuous at high packing fractions \cite{BJ09}.   Over the whole range explored in our experiments Eq.~\ref{eqn:HBmodel_plus} fits the data well, as demonstrated by the example in Fig.~\ref{fig:HBfit}. 

The lower boundary of the shear thickening region occurs at the stress $\tau_m$ and shear rate $\dot\gamma_m$ corresponding to the local viscosity minimum.  Differentiating $\eta\equiv \tau/\dot\gamma$  and eliminating $a_2$ via Eq.~\ref{eqn:HBmodel_plus} evaluated at $\tau_m$ gives

\be
\tau_m = \tau_{HB}(\dot\gamma_m) + \frac{\epsilon}{2(1-\epsilon)}\left[ \tau_{HB}(\dot\gamma_m) + \tau_y\right] \ .  
\label{eqn:tau_min}
\ee

\noindent Eq.~\ref{eqn:tau_min} is in a form that directly shows how the shrinkage of the shear thickening regime depends on the shear thinning terms.  The model parameters $\epsilon$, $\tau_y$ and $a_1$ are obtained by fitting the data to Eq.~\ref{eqn:HBmodel_plus} for each value of $B$, $E$, and $\phi$.  The values of $a_2$ and $\epsilon$ are found to be independent of $B$ and $E$ (see Suppl. Mat.), which can be seen from the overlay of shear thickening curves at higher stresses in Fig.~\ref{fig:STwfield}, indicating that the shear thickening stress is independent of attractions.  This is in contrast to weaker, `continuous' shear thickening due to hydrodynamics where attractions were found to affect the shear thickening stress \cite{GZ04}.   Equation \ref{eqn:tau_min} is evaluated at the measured $\dot\gamma_m$, $\tau_y$, and $a_1$ for each $B$, $E$, and $\phi$, and a fixed value of $\epsilon$ for each panel, as shown by the solid red symbols in Fig.~\ref{fig:phasediagrams}.  This describes the lower phase boundary very well, typically within 12\%.  We note that Eq.~\ref{eqn:tau_min} along with an equation for $\dot\gamma_m$ (obtained from Eqs.~\ref{eqn:HBmodel}-\ref{eqn:tau_min}, see Suppl. Mat.) can also be used to predict the phase boundary with attractions without measuring $\dot\gamma_{m}$ for all field values, assuming only that the shear thickening stress is independent of the shear thinning mechanism.  

The agreement of Eq.~\ref{eqn:tau_min} with the measured phase boundaries in Fig.~\ref{fig:phasediagrams} demonstrates that the lower shear thickening phase boundary is set by the mechanism that produce shear thinning.  Because of the second term in Eq.~\ref{eqn:HBmodel}, this is true whether or not the shear thinning includes a yield stress.  It is also independent of whether the shear thickening stress term is affected by the parameter that controls shear thinning if the shear thickening and shear thinning terms add linearly as seen in Fig.~\ref{fig:phasediagrams}c and for 'continuous' shear thickening \cite{SWB03,GZ04}.   The fact that this model reproduces the measured phase boundary confirms that, for 'discontinuous' shear thickening, the effect of attractions is to increase the shear thinning stress which hides shear thickening, rather than to affect the shear thickening stress directly.  When $B$, $E$, or $\phi$ become large enough that shear thickening is not observed, $\tau_m$ (open red symbols in Fig.~\ref{fig:phasediagrams}) becomes higher than the shear thickening stress range.  In the limit of $\epsilon=0$, Eq.~\ref{eqn:tau_min} reduces to $\tau_m=\tau_{HB}(\dot\gamma_m)$, so the stress at the phase boundary becomes equal to the shear thinning stress.  Thus, the shear thickening regime starts to shrink when the shear thinning stress exceeds the stress at the onset of shear thickening, and it is eliminated when the shear thinning stress exceeds the stress at the viscosity maximum.  Regardless of the particulars of the mathematical model, this is a good approximation as long as there is a sharp upturn in $\tau(\dot\gamma)$ which is the defining feature of 'discontinuous' shear thickening, and the shear thickening stress term is independent of the shear thinning mechanism.  This interpretation remains true for $\epsilon>0$ with corrections according to Eq.~\ref{eqn:tau_min}.  It also holds in cases where there is a Newtonian regime before the onset of shear thickening, regardless of the value of $\epsilon$ (see Suppl. Mat.).

Our simple model predicts the shear thickening phase boundaries without knowing detailed particle properties or microstructure.  As long as the shear thinning mechanism produces a stress term that adds linearly to the shear thickening stress term, all sources of shear thinning have the same effect of hiding shear thickening, regardless of the mechanisms that produces shear thickening.  This description in terms of stress scales is not dependent on size scale and, in principle, might be applicable also to 'discontinuous' shear thickening in colloidal (i.e. Brownian) systems.  In colloids, however, different mechanisms for shear thickening and thinning might become relevant, for example a shear thinning stress term due to Brownian motion \cite{BBV02}.  
 
The existence of an upper threshold beyond which shear thinning mechanisms will overwhelm shear thickening explains why in most cases attractions completely eliminate shear thickening \cite{Ba89} while for some fluids with weak interparticle attractions it has been reported to exist  \cite{OKW08, LDH03}.  In typical suspensions, attractions are often due to particle-fluid surface tension.  An example is the common observation that cornstarch (a hydrophilic particle) shear thickens in water but not in hydrophobic liquids \cite{PJ41}.   One can then ask if all suspensions will shear thicken once the shear thinning stresses are small.  In the experiments reported here on a variety of suspensions consisting of particles including cornstarch, glass, and PEG, in a variety of fluids with different density matching, modified surface properties, roughness, shapes, and measuring conditions, we always observed 'discontinuous' shear thickening.   Including a variety of other suspensions we studied, we found no examples where the shear thinning stress was small (less than the order of 5 Pa for particles on the order of 10-100 $\mu$m) that did not shear thicken at near-sedimentation packing fractions.   Inductively this suggests the phenomenon of 'discontinuous' shear thickening is general to all hard particle suspensions at near-sedimentation packing fractions provided that the shear thinning stresses are below a threshold \cite{BB85, Ba89}.  

The combination of ER or MR effects with shear thickening opens up possibilities for the design of field-responsive shear thickening fluids in dampers or impact absorbers  \cite{Lord, SWB03, ZWG08}.  We note that earlier suggestions presumed that the applied fields would control the critical shear rate \cite{Lord}, but this is only true for weaker shear thickening ($\epsilon>0$), where both the critical stress and shear rate vary with field.   In the limit where shear thickening becomes discontinuous ($\epsilon=0$), our findings show that the critical shear rate is controlled by the particle packing fraction \cite{BJ09}, while the critical stress can be tuned either passively with particle-fluid chemistry or actively with fields.

\subsection{Acknowledgements}

This work was supported by DARPA through Army grant W911NF-08-1-0209.  EB acknowledges additional support by the NSF MRSEC program under DMR-0820054.  We thank J. Xu for performing the optical tweezer measurements, K. Herlihy and J. Nunes for help with the magnetite-containing particle synthesis, L. Mair and R. Superfine for assistance with optical microscope images in calibrated magnetic fields, and J. Sprague and M. Hunter  for  assistance with manufacturing of PRINT particles.  

\subsection{Author Contributions}

E.B. and H.M.J. conceived of the study and wrote the manuscript.  All team members were involved in conception of manufactured particles that show both a magnetorheological and shear thickening effect.   H.Z, N.A.F, D.E.B., and J.M.D. were responsible for design and initial fabrication of these particles.   N.A.F, B.W.M, and J.M.D were responsible for production of gram quantities of these particles.  E.B. and C.S.O. were responsible for the rheological measurements.  E.B. analyzed the data.

\section{Supplementary Information}

\subsection{Additional experimental details}
\label{sec:experimentaldetails}

\begin{figure}                                                
\centerline{\includegraphics[width=2.75in]{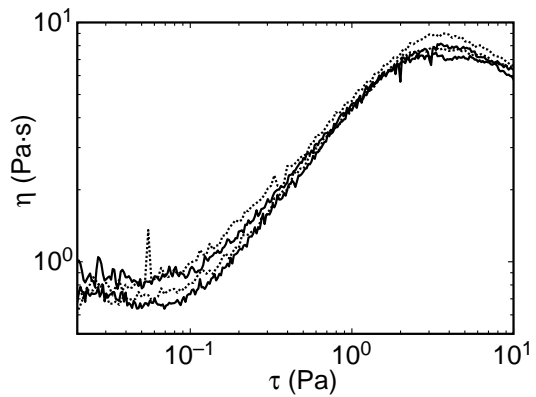}}
\caption{Viscosity curves showing an example of hysteresis and ramp rate dependence for 100 $\mu$m glass spheres in mineral oil at $\phi=0.552$.  Solid line: 100 measurement points per decade of stress, 10 s per point.  Dotted line:  50 measurement points per decade of stress, 10 s per point.  Upper curves correspond to a decreasing stress ramp.  Lower curves correspond to an increasing stress ramp.}
\label{fig:hysteresis}                                        
\end{figure}

In the rheometer the torque $T$ on the tool and its rotation $\omega$ rate were measured and converted to a shear stress $\tau$ and a shear rate  $\dot\gamma $.  Both parallel plate and Couette geometries were used (for the parallel plate geometry $\tau = 2T/(\pi R^3)$ and  $\dot\gamma = R\omega/d$, where $R$ is the plate radius and $d$ the gap size).   The shear stress and shear rate describe the mechanical response in a geometry-independent form, but we do not imply or require a linear flow profile.   Data for Figs.~1\label{fig:hydrophobic} and 3\label{fig:confinement} were taken with increasing controlled stress to resolve the steep shear thickening, while data for Fig.~2\label{fig:STwfield} were taken with controlled shear rate to allow for a better fit of  the Herschel-Bulkley model, Eq.~1\label{eqn:HBmodel}.  Care was taken that no fluid extended outside the parallel plates and the particles were confined to the space between the plates by surface tension.  Samples were pre-sheared for 200 s to stresses above the shear thickening region immediately before experiments commenced after which measurements were found to be reproducible within a typical variation of about 10-20\%.  Measurements reported were mostly taken at ramp rates of 500 s per decade of controlled stress or shear rate.  Increasing as well as decreasing ramps with different ramp rates were used to check for hysteresis, thixotropy, and transients.   An example comparison of ramp rate and direction dependence is shown in Fig.~\ref{fig:hysteresis}.  Some ramp-rate independent hysteresis was observed between increasing and decreasing stress measurements.  While the magnitude varied from suspension to suspension and typically  about 20\% of the viscosity in the shear thickening regime, the curves were never qualitatively different. 

 For clarity, the data shown are for one ramp direction.  We checked for reversibility by shearing suspensions in the shear thickening regime and then immediately ceasing shear;  the result was that the stress relaxes to the zero-shear limit within seconds.  Different gap sizes between 0.5-1 mm  were used to check for finite size effects.  Reported experiments were done with smooth plates.   Rough plates were also used in some experiments to check for slip and no significant differences were found.  Reported packing fractions $\phi$ are based on measured particle and fluid quantities mixed together before shearing.  

To observe the discontinuous viscosity curves as $\epsilon \rightarrow 0$, confinement of the sample is important.   This can be achieved by either using a Couette geometry or avoiding slop in a parallel plate geometry \cite{BJ09sup, FHBOB08sup}.  Non-density matched samples were measured in a parallel plate geometry to minimize the weight on the packing which produces a yield stress in a Couette geometry \cite{FBOB09sup}.  Attempts to measure shear thickening of glass spheres in mineral oil in a Couette geometry resulted in a large yield stress due to this sedimentation and no shear thickening.  This observation is perfectly consistent with our conclusion that a yield stress from any source can hide shear thickening.  

The glass spheres were obtained from MoSci corporation (Class IV).  They were sieved through mesh sizes -120+170 and were measured to have a mean diameter of 89 $\mu$m with a standard deviation of 12 $\mu$m.   For the surface tension experiments, the glass spheres were mixed into water after the surfactant so the total fluid volume matched that of the case without surfactant.  This ensured that the surfactant diffused throughout the sample and the volume fraction did not vary between the two experiments.  These measurements  were done with a parallel plate setup with a 50 mm diameter rotating top plate with a 0.83 mm gap. 

Cornstarch was chosen as a prototypical shear thickener for the packing fraction dependent experiments.  Argo cornstarch was used at ambient conditions of 23$^{\circ}$C and 42\% humidity which included some water weight.  The suspensions were density matched for the reported experiments so that the yield stress in this case was due to confinement \cite{FBOB09sup}.  For the cornstarch data shown we used a Couette geometry consisting of a 26.6 mm diameter cylinder in a cup with a gap of 1.13 mm.  We found that mismatching the density for starch did not affect the critical packing fraction because in this case the hydrostatic pressure from the weight of the packing is still much lower than the yield stress above the critical packing fraction.   The same behavior was also obtained in a parallel plate geometry.

For electrorheology measurements, any dielectric particle in a non-conducting fluid will work. We used hydrophobically-coated glass in mineral oil so the particle-fluid surface tension was minimized.  To apply the dc electric field, 50 mm diameter parallel metal plates with a gap of 0.88 mm were used as electrodes.  The reported electric field value is the applied voltage divided by the gap size.  The rotating upper plate fixture made electrical contact via a wire brush. This added a constant friction corresponding to about 0.1 Pa which limited the stress resolution of those measurements.  After subtracting this offset, the data in the limit of vanishing applied field matched the zero-field value measured without using the wire brush.  Therefore, the stress resolution limit did not artificially set the measured onset of shear thickening.

Most magnetorheological fluids have a yield stress even in the absence of a field.  To obtain a sample that showed both shear thickening and a magnetorheological response we engineered a suspension to minimize particle-fluid surface tension with particles that could be filled with magnetic material.  To achieve this we used the PRINT$^{\textregistered}$ process.  Typically, the monomer solution was prepared as follows: 0.30 g of magnetite (black iron oxide, average particle diameter = 0.2 $\mu$m, Polysciences, Inc.), 0.02 g of 1-hydroxycyclohexyl phenyl ketone (HCPK, Aldrich), and 0.01 g of fluorescein o-acrylate (Aldrich) were placed into an Eppendorf tube followed by the addition of 0.1 ml of N,N-dimethylformamide (DMF, Aldrich). The monomer mixture was then mixed thoroughly by vortex mixing to dissolve the HCPK photoinitiator and the fluorescein o-acrylate fluorophore. Lastly, 0.67 g of ethoxylated(20) trimethylolpropane triacrylate (MW = 1176 g/mol, SR415, Sartomer) was added to the monomer mixture and vortex mixed again. The resulting solution was composed of 30\% (w/w) magnetite, 67\% (w/w) triacrylate, 2\% (w/w) HCPK, and 1\% (w/w) fluorescein o-acrylate. The rod-shaped particles were then fabricated using the PRINT process, which has been described elsewhere \cite{RMEEDD05sup, HND08sup}. Molds for fabrication of PRINT particles were supplied by Liquidia Technologies.  For the magnetorheological experiments, the particles were suspended in poly(ethylene glycol) dimethyl ether (Mn = 500 g/mol, Aldrich). These measurements were conducted in a 20 mm diameter parallel plate geometry with a gap of 0.9 mm. For imaging purposes, DyLight 549 Maleimide (MW = 1007 g/mol, Fisher) was used as the fluorophore.  

\subsection{Yield stress}

The yield stress can be defined differently and thus measured in several different ways.   A static or dynamic yield stress can be measured for either increasing or decreasing control ramps, respectively, and each can be done with either controlled stress or shear rate.  In our experiments, each method yielded similar yield stress values.  Some hysteresis was observed between the static and dynamic yield stresses, which was larger for faster ramp rates.  At slower ramp rates the hysteresis loops converged to a relatively small difference (less than a factor of 2).  The reported data were taken at ramp rates in this latter regime.

By defining the viscosity as $\eta\equiv \tau/\dot\gamma$, it is infinite below the yield stress since the shear rate is zero.   Given that shear thickening requires the viscosity to increase with stress, $\eta$ must first drop to finite values, so a viscosity function with a continuous first derivative necessarily displays shear thinning before entering a shear thickening region.   A different value for the yield stress does not change this general behavior, but can move the onset of the shear thickening region. The shear thickening stress range can therefore depend somewhat on the yield stress definition, specifically if the lower shear thinning region is small.  For the purposes of comparing shear thinning and shear thickening stresses to determine the shear thickening regime, the particular criterion for evaluating the yield stress is irrelevant as long as it is done consistently.  The conclusion that the shear thickening phase boundary is determined by the shear thinning stress does not depend on which yield stress is measured or any specific form for the model.  We chose the Herschel-Bulkley model with exponent 1/2 only because it fits the data well (but see below for other exponents).  

\subsection{Connection between particle interactions and macroscale rheology}

The connection between field-induced interparticle attractive forces and the yield stress can be explained through electrorheology models \cite{Zu93sup}.  In an applied electric field $E$ the induced dipole moment density scales as $\beta\epsilon_0 E$ where $\beta$ is an effective dielectric constant  that depends on particle and fluid dielectric constants and saturates at values of order unity for all but a near-exact dielectric match.   The resulting net force between neighboring particles scales as $F \sim \epsilon_0 \epsilon_L \beta^2 E^2 a^2$ for particle diameter $a$ and liquid dielectric constant $\epsilon_L$.  The yield stress can be obtained by dividing this attractive force by an effective particle surface area, giving a yield stress scale $\tau_{y} = 12\pi \epsilon_0\epsilon_L\beta^2 E^2$ \cite{Zu93sup}.  This result is shown in Fig.~2\label{fig:STwfield}b and agrees with the measured yield stress at high field values.  Settling becomes more important below the gravitational stress scale $\Delta\rho g a \approx 1.5$ Pa where $\Delta\rho$ is the density difference and $g$ is the acceleration of gravity.  If the attractive stress is below this scale, particles will tend to settle at lower shear stresses rather than form chains to jam the system.  This is likely the reason that the yield stress falls below the $E^2$ scaling at lower field values.  The agreement of the yield stress with the attractive stress scale shows that the yield stress and hence the shear thickening regime can be connected to individual particle properties.  It is interesting to note that this calculation did not require any specific knowledge of the flow or packing structure.   

The yield stress scale could be put in terms of the attractive force divided by particle surface area, or equivalently the attractive energy per unit volume.  This attractive stress scale can be calculated for other types of attractions as well to relate the yield stress to microscopic properties.  For example, we can estimate the expected yield stress from other sources of interactions that might be operative between cornstarch particles.   To check for this, we used optical tweezers to place two cornstarch particles next to each other in water and allowed them to diffuse.  An attractive or repulsive potential can be measured by observing the probability distribution of the separation distance over time.  In the tweezer  experiment the resolution was about 1 pN and down to this instrumental limit no attractive or repulsive forces were observed.  Dividing this value by particle surface area puts the upper limit on the yield stress due to attractions at around $10^{-2}$ Pa.  This is consistent with the fact that we did not observe any yield stress in cornstarch suspensions down to our instrument resolution of $10^{-3}$ Pa at low packing fractions.

\subsection{Approximations of the phase boundary}
\label{sec:fitting}

\begin{figure}                                                
\centerline{\includegraphics[width=2.in]{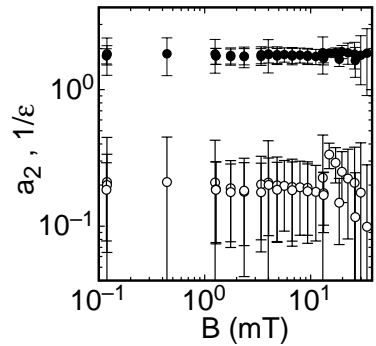}}
\caption{  Parameter values obtained from fitting Eq.~2 to the data on shear thickening with the MR effect.  ($\circ$):  $a_2$ in units corresponding to Pa for stress and Hz for shear rate.  ($\bullet$): $1/\epsilon$.  No resolvable trend in either parameter is found as $B$ is varied.  }  
\label{fig:shearthickenfit}                                        
\end{figure}

To understand how the lower shear thickening phase boundary is determined by the shear thinning stress, we next discuss various approximate solutions based on Eq.~3\label{eqn:tau_min}.  The basic idea is as follows: Given that the shear thickening stress is independent of the strength of attractions (see Fig.~2\label{fig:STwfield}) and described by Eq.~2\label{eqn:HBmodel_plus}, the phase boundary can be determined by measuring $\dot\gamma_m$, $a_2$ and $\epsilon$ for zero field, and then calculating $\tau_y(B)$ and $a_1(B)$ as a function of applied field.  To check the feasibility of this approach, we first show that the fit parameters $a_2$ and $\epsilon$ are independent of the applied field.  This is done by fitting the experimental data to Eq.~2\label{eqn:HBmodel_plus} for various field values, covering a range up to 80\% of the shear rate at the viscosity maximum or up to 3 Hz if there is no maximum (this covers roughly the same shear rate range).  The result, shown in Fig.~\ref{fig:shearthickenfit}, confirms the claim that the shear thickening stress is unaffected by (field-induced) attractions. For brevity, we will show the subsequent analysis only for the magnetorheology data but it applies for the electrorheology data as well (the $\phi$ dependence is more complicated because the packing fraction controls the shear thickening behavior itself in addition to tuning the yield stress). 

\begin{figure}                                                
\centerline{\includegraphics[width=2.75in]{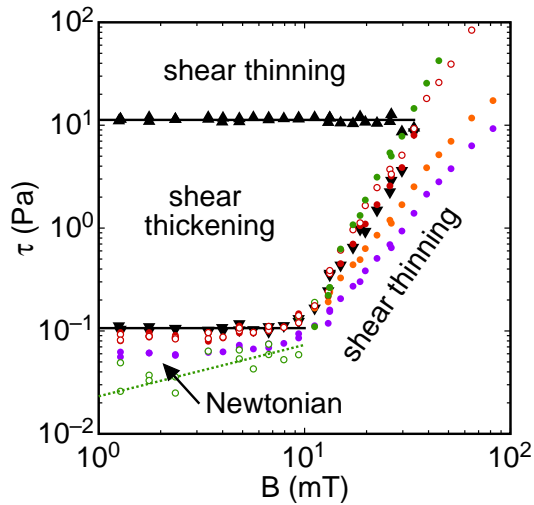}}
\caption{Phase diagram showing different approximations of the shear thickening phase boundary. Lower ($\blacktriangledown$) and upper ($\blacktriangle$) boundaries of the shear thickening regime are as in Fig.~4\label{fig:phasediagrams}a.   Purple circles ({\color{Purple} $\bullet$}): estimate in the limit of $\epsilon=0$ giving $\tau_m = \tau_{HB}(\dot\gamma_m)$ evaluated at $\dot\gamma_{m,0}$.  Orange circles ({\color{Orange} $\bullet$}): estimate accounting for the non-zero $\epsilon$ using the measured value of $\epsilon=0.55$ in Eq.~3\label{eqn:tau_min}.   Open red circles ({\color{Red} $\circ$}): estimate further accounting for the change in $\dot\gamma_m$ with $\epsilon$ and $B$ by evaluating Eq.~3\label{eqn:tau_min} at $\dot\gamma_m$ calculated from Eq.~\ref{eqn:dotgammamin}.  Solid red circles ({\color{Red} $\bullet$}):   Eq.~3\label{eqn:tau_min} at the measured $\dot\gamma_m$.  Solid green circles ({\color{Green} $\bullet$}): prediction from Eq.~3\label{eqn:tau_min} at $\dot\gamma_m$ calculated from Eq.~\ref{eqn:dotgammamin} using only data obtained at $B=0$ and the lower shear thinning regime for larger $B$.  Open green circles  ({\color{Green} $\circ$}):  evaluating Eq.~3\label{eqn:tau_min} at $\dot\gamma_m$ calculated from Eq.~\ref{eqn:dotgammaminfit} with $\alpha=1/2$.  Dotted green line: fit of the open green circles indicating the phase boundary between shear thinning and Newtonian regimes.
}  
\label{fig:phasediagramapprox}                                        
\end{figure}

The simplest approximation of the phase boundary is to assume the discontinuous limit $\epsilon=0$.  When varying the yield stress via attractive interactions, the fact that the shear thickening part of $\tau(\dot\gamma)$ is steep means the  the onset of shear thickening occurs at a nearly constant $\dot\gamma_m$.  This allows for a simplification since we can approximate $\dot\gamma_m$ in Eq.~3\label{eqn:tau_min} by its value measured for zero applied field.  This gives $\tau_m = \tau_{HB}(\dot\gamma_m)$ indicating that the stress at the phase boundary is equal to the shear thinning stress.   We evaluate Eq.~1\label{eqn:HBmodel} using fit values of $\tau_y(B)$ and $a_1(B)$ for each applied field evaluated at $\dot\gamma_{m,0}$, the measured onset at zero applied field.  This is shown as purple symbols in Fig.~\ref{fig:phasediagramapprox}, along with the data from Fig.~4\label{fig:phasediagrams}a.  This $\epsilon=0$ approximation underestimates the onset of shear thickening by  $50\pm20$\% (errors indicate a standard deviation).  The fact that this approximation gives the threshold where attractions begin to move the onset and the increase in the onset with field within about a factor of 2 confirms that the phase boundary is determined by the shear thinning stress.   

A better quantitative match to the lower phase boundary can be obtained by accounting for the non-zero $\epsilon$.  The orange symbols in Fig.~\ref{fig:phasediagramapprox} are plotted for the same $\tau_y$ and $a_1$ as before but now using the measured value of $\epsilon=0.55$.  This increases the predicted $\tau_m$ by a factor of 1.5.  As a result, the measured boundary is underestimated by $9\pm 9$\%  in the low-field region and at higher fields, where attractions are reducing the shear thickening regime, by $33\pm 20\%$.  Overall, this better predicts the point where the attractions are strong enough to increase the onset of shear thickening but still underestimates the effect of attractions.  

The next correction is to account for the change in $\dot\gamma_m$ with attractions for $\epsilon > 0$.  Using techniques similar to those used in Sec.~\ref{sec:derivation}, an exact implicit equation can be written for $\dot\gamma_m$ in a form that shows how $\dot\gamma_{m}$ varies with $\epsilon$:

\be
\dot\gamma_m(B)^{\frac{2-\epsilon}{2\epsilon}} = \left(\dot\gamma_{m,0}\right)^{\frac{2-\epsilon}{2\epsilon}}  
+ \frac{\epsilon}{2(1-\epsilon)a_2} \left[\Delta a_1(B)+\frac{2\tau_y(B)}{\sqrt{\dot\gamma_m(B)}} \right] 
\label{eqn:dotgammamin}
\ee

\noindent  where $\Delta a_1(B) \equiv a_1(B)-a_1(0)$.  Eq.~\ref{eqn:dotgammamin} reduces to $\dot\gamma_m = \dot\gamma_{m,0}$ for no attractions [$\Delta a_1(B)=0$, $\tau_y(B) = 0$] as expected or for step-function stress-shear rate curves ($\epsilon=0$) as already claimed, and it justifies the simplification $\tau_m = \tau_{HB}(\dot\gamma_{m,0})$ in the limit of $\epsilon=0$.  Such constant onset shear rate when the yield stress is varied in the limit of $\epsilon=0$ contrasts with the constant onset stress when the packing fraction is varied (not including the contribution of the yield stress) \cite{BJ09sup, MW01asup, EW05sup}.  

\begin{figure}                                                
\centerline{\includegraphics[width=2.in]{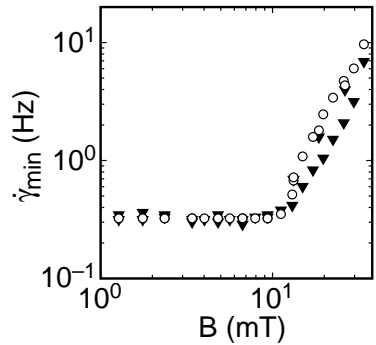}}
\caption{   Solid triangles ($\blacktriangledown$): measured shear rate at the onset of shear thickening $\dot\gamma_m$.   Open cirlces ($\circ$): calculated $\dot\gamma_m$ from Eq.~\ref{eqn:dotgammamin}.   }  
\label{fig:sratemin}                                        
\end{figure}

Since Eq.~\ref{eqn:dotgammamin} is an implicit equation, it must be evaluated numerically. We note that since the yield stress is a small contribution to the overall shear thinning stress as seen in Fig.~4\label{fig:phasediagrams}a, i.~e.~$a_1\dot\gamma_m^{1/2} \gg \tau_y$, the rightmost term with $\dot\gamma_m(B)$ in the denominator is small.  Starting with the value of $\dot\gamma_m(B) = \dot\gamma_{m,0}$ on the right hand side, we can evaluate Eq.~\ref{eqn:dotgammamin} iteratively. The value of $\dot\gamma_m$ converges to within a few percent after only 2 iterations.  Thus for a simple explicit estimate one can set $\dot\gamma_m(B) = \dot\gamma_{m,0}$ on the right side.  This estimate of $\dot\gamma_m$ is shown in Fig.~\ref{fig:sratemin} in comparison to the measured $\dot\gamma_m$.  It is seen that the model captures the increase in $\dot\gamma_{min}$ with attractions, particularly the point where attractions start to increase $\dot\gamma_m$ which occurs at the same point the stress starts to increase due to the yield stress pushing up the onset of shear thickening.  Beyond that point the model overestimates the measured values by $30\pm 20\%$. 

We now evaluate Eq.~3\label{eqn:tau_min} using the calculated value of $\dot\gamma_m(B)$ from Eq.~\ref{eqn:dotgammamin}, $\epsilon=0.55$, and the fit values of $\tau_y(B)$ and $a_1(B)$.  This is shown as the open red symbols in Fig.~\ref{fig:phasediagramapprox} (same as in Fig.~4\label{fig:phasediagrams}a) and gives the entire phase boundary within a standard deviation of 28\%.   For the packing fraction dependence in Fig.~4\label{fig:phasediagrams}c, there is no comparable prediction for the onset shear rate  because the shear thickening term varies with packing fraction.  Thus the open red symbols in panel c correspond to Eq.~3\label{eqn:tau_min} evaluated at the smallest measured onset shear rate.  

A check on the validity of Eq.~3\label{eqn:tau_min} for describing the phase boundary can be made by using the fit parameters $\tau_y(B)$ and $a_1(B)$ and $\dot\gamma_m(B)$ measured at the minimum of $\eta(\tau)$.  This is shown as solid red symbols in Fig.~\ref{fig:phasediagramapprox} (same as in Fig.~4\label{fig:phasediagrams}a) which agrees with the measured phase boundary to within a standard deviation of 12\%.  For comparison, if we repeat measurements keeping all control parameters constant, the typical variation in the measured $\tau_m$ is 11\%.  Thus, the model is accurate in describing the onset of shear thickening up to the resolution of the data.

Given the assumptions that the shear thinning and thickening terms add linearly and the shear thickening term is independent of attractions, in principle we can predict the phase boundary using only the shear thickening curve at zero field and the effect of attractions on $\tau_{HB}$.  The above analysis was all done using fits of data up into the shear thickening regime.  To show the predictive power of the model, we now perform the fit to the data for zero field only to obtain $a_2$, $\epsilon$, and $\dot\gamma_{m,0}$ and keep these fixed.  We then fit Eq.~2\label{eqn:HBmodel_plus} to data for non-zero field only up to some cut-off $\dot\gamma < 0.3$ which is in the lower shear thinning region for data with non-zero field.  This allows us to obtain $\tau_y$ and $a_1$.   The fitting cutoff can be chosen based on the zero-field data because the attractions always increase the onset of shear thickening.  We then evaluate Eq.~3\label{eqn:tau_min} using the fit values of $\tau_y$ and $a_1$, and $\dot\gamma_m$ from Eq.~\ref{eqn:dotgammamin}.  This gives the solid green symbols shown in Fig.~\ref{fig:phasediagramapprox}a and b.   

This prediction of the phase boundary without using any data from the shear thickening regime except at zero field overestimates the  phase boundary by $20\pm60$\%.  While this agreement is not as good as when we fit the full data set, it shows that the effect of attractions on shear thickening can be predicted within about a factor of two.  In the case where packing fraction is varied, the shear thickening stress itself changes; $a_2$ and $\epsilon$ vary with $\phi$, which shifts $\dot\gamma_m$ without the influence of attractions.  Thus either $a_2$ and $\epsilon$, or $\dot\gamma_m$ must be obtained as a function of $\phi$ to determine the phase boundary.

A Newtonian regime is sometimes found before the onset of shear thickening, for example in Fig.~2\label{fig:STwfield}a.  We did not explicitly include a Newtonian term in the model.  However, the generalized expression for $\tau_m$ in Eq.~\ref{eqn:tau_min_general} can apply for a Newtonian stress term when $\alpha =1$ (see the next Section, below).
The onset of shear thickening can still be expressed by $\tau_m$ in Eq.~\ref{eqn:tau_min_general} if a Newtonian stress term is added to Eq.~2\label{eqn:HBmodel_plus}, although the value of $\dot\gamma_{m,0}$  would generally increase.   We expressed Eq.~\ref{eqn:dotgammamin} in a perturbative form rather than as a simple dependence on the fit parameters so it still applies in the case where there is a Newtonian regime whether or not it can be described by a linear addition to the stress-shear rate relation of Eq.~2.  If instead Eq.~\ref{eqn:dotgammaminfit} is used for $\dot\gamma_m$ without consideration of a Newtonian regime, the phase boundary would be underestimated for weak attractions (open green symbols in Fig.~\ref{fig:phasediagramapprox}).  Thus the difference between the open red and green circles in Fig.~\ref{fig:phasediagramapprox} is due to the Newtonian regime.  

For simplicity we omitted any Newtonian regime from the main paper and included it in the shear thinning regime.  This does not change the conclusions but to be more general we can restate them in a way that includes the possibility of a Newtonian regime.  This regime disappears for stronger attractions when the shear thinning stress overwhelms the Newtonian stress term.    Thus for shear thickening to occur in general, the shear thickening stress must overcome the sum of shear thinning and Newtonian stresses.  On the other hand, for attractions to affect the onset of shear thickening, they must exceed a threshold equal to the inherent shear thinning and Newtonian stresses at the onset.

\subsection{Notes on possible mechanisms for shear thickening}

The phenomenological approach presented here does not address the microscopic origin of shear thickening in suspensions, but the data put constraints on the region of validity for existing models.  While hydrodynamic models have successfully described Continuous Shear Thickening which occurs at lower packing fractions and higher shear rates, they have not been able to reproduce the steep stress/shear-rate relation ($\epsilon \approx 0$) characteristic of Discontinuous Shear Thickening, instead the smallest value of $\epsilon$ allowed in those models is $1/2$  \cite{BBV02sup}.   Inertial granular models have a similar limitation \cite{Ba54sup}.  In addition, hydrodynamic models predict that the shear thickening stress should be affected by attractions \cite{SWB03sup, GZ04sup}. However, for Discontinuous shear thickening this is not what we observe.

\subsection{Derivation of Eq.~3\label{eqn:tau_min}}
\label{sec:derivation}

Here we derive an expression for the onset of shear thickening from $\tau(\dot\gamma)$ given in Eq.~2\label{eqn:HBmodel_plus} with a generalized Herschel-Bulkley form for the shear thinning term $\tau_{HB} = \tau_y+a_1\dot\gamma^{\alpha}$ .   The onset corresponds to the local viscosity minimum which satisfies 

\be
0 = \frac{d\eta}{d\dot\gamma}\bigg|_{\dot\gamma_m} = -\frac{\tau_y}{\dot\gamma_m^2} + (\alpha-1)a_1\dot\gamma_m^{\alpha-2} + (1/\epsilon-1)a_2 \dot\gamma_m^{1/\epsilon-2} \ .
\label{eqn:derivativeatonset}
\ee

\noindent  Rearrangement gives

\be
a_2\dot\gamma_m^{1/\epsilon} = \frac{\epsilon}{1-\epsilon}\left[ \tau_y + (1-\alpha)a_1\dot\gamma_m^{\alpha}\right] \ .
\label{eqn:taumin_condition}
\ee

\noindent Substituting Eq.~\ref{eqn:taumin_condition} into Eq.~2 evaluated at $\dot\gamma_m$ gives

\be
\tau_m = \tau_{HB}(\dot\gamma_m) + \frac{\epsilon}{1-\epsilon}\left[\tau_y+(1-\alpha)a_1\dot\gamma_m^{\alpha}\right] \ .
\label{eqn:tau_min_general}
\ee

\noindent  Setting $\alpha=1/2$ gives Eq.~3.  This shows that in the limit of $\epsilon=0$ the onset of shear thickening is equal to $\tau_{HB}(\dot\gamma_m)$ regardless of the form of the shear thinning term.

\subsection{Derivation of Eq.~\ref{eqn:dotgammamin}}

Here we derive the expression for the shear rate at the onset of shear thickening $\dot\gamma_m$, similar to the derivation for $\tau_m(\dot\gamma_m)$. We rearrange Eq.~\ref{eqn:derivativeatonset} to get

\be
 \dot\gamma_m^{\frac{1}{\epsilon}-\alpha} =  \frac{\epsilon}{(1-\epsilon)a_2} \left[(1-\alpha)a_1 + \tau_y \dot\gamma_m^{-\alpha}\right] \ .
 \label{eqn:dotgammaminfit}
\ee

\noindent To put this in the form of Eq.~\ref{eqn:dotgammamin} to directly describe the perturbation in $\dot\gamma_m$ with an additional shear thinning term, we evaluate this for zero additional attractions to obtain

\be
 \dot\gamma_{m,0}^{\frac{1}{\epsilon}-\alpha} =  \frac{\epsilon}{(1-\epsilon)a_2} (1-\alpha)\left(a_1-\Delta a_1 \right)
\ee

\noindent where $\Delta a_1 \equiv a_1 - a_{1,0}$ and $a_{1,0}$ is the value of $a_1$ for the unperturbed state without the additional shear thinning term.  Substituting this back in to Eq.~\ref{eqn:dotgammaminfit} to eliminate $a_1$ results in 

\be
 \dot\gamma_{m}^{\frac{1}{\epsilon}-\alpha}  =  \dot\gamma_{m,0}^{\frac{1}{\epsilon}-\alpha} +  \frac{\epsilon}{(1-\epsilon)a_2} \left[(1-\alpha)\Delta a_1 + \tau_y\gamma_m^{-\alpha} \right]   \ .
\ee

\noindent We evaluate this at $\alpha=1/2$ to obtain Eq.~\ref{eqn:dotgammamin}.

\end{document}